\documentclass[amsmath,amssymb]{nature}

\usepackage{graphicx}
\usepackage{amsmath}
\usepackage{amssymb}
\usepackage{ bbold }
\usepackage{ bm }

\newcommand{\onlinecite}[1]{\nocite{#1}\citenum{#1}}

\def\nm{{\ {\rm nm}}}                       
\def\micron{{\ \mu{\rm m}}}                 

\def\mT{{\ {\rm mT}}}                       

\def\Hz{{\ {\rm Hz}}}                       

\def\us{{\ \mu{\rm s}}}                     
\def\ms{{\ {\rm ms}}}                       

\def\Er{E_R}                            
\def\kr{k_R}                            
\def\Rb87{^{87}\text{Rb}}                     
\def\Na23{^{23}\text{Na}}                     
\def\K41{^{41}\text{K}}                     
\def\Li6{^{6}\text{Li}}                       
\def\Li7{^{7}\text{Li}}                       

\DeclareMathAlphabet\mathbfcal{OMS}{cmsy}{b}{n}

\def\ket#1{\mathinner{|{#1}\rangle}}

{\catcode`\|=\active
  \gdef\Braket#1{\left<\mathcode`\|"8000\let|\BraVert {#1}\right>}}
\def\BraVert{\egroup\,\mid@vertical\,\bgroup}

\DeclareMathAlphabet{\mathpzc}{OT1}{pzc}{m}{it}

\title{The spin Hall effect in a quantum gas}

\author{M. C. Beeler$^1$, R. A. Williams$^1$, K. Jim\'enez-Garc\'ia$^{1, 2, 3}$, L. J. LeBlanc$^1$, A. R. Perry$^1$, \& I. B. Spielman$^1$}

\begin{document}

\newcommand{\ex}{${\bf e}_x$}
\newcommand{\ey}{${\bf e}_y$}
\newcommand{\ez}{${\bf e}_z$}

\maketitle

\begin{affiliations}
 \item Joint Quantum Institute, National Institute of Standards and Technology and University of Maryland, Gaithersburg, MD 20899, USA
 \item Departamento de Fisica, Centro de Investigaci\'{o}n y Estudios Avanzados del Instituto Polit\'{e}cnico Nacional, Mexico D.F. 07360, Mexico
 \item Current Address: The James Franck Institute and Department of Physics, The University of Chicago, Chicago, IL 60637, USA
\end{affiliations}

\begin{abstract}

Electronic properties like current flow are generally independent of the electron's spin angular momentum, an internal degree of freedom present in quantum particles. The spin Hall effects (SHEs), first proposed 40 years ago\cite{Perel1971}, are an unusual class of phenomena where flowing particles experience orthogonally directed spin-dependent Lorentz-like forces, analogous to the conventional Lorentz force for the Hall effect, but opposite in sign for two spin states.  Such spin Hall effects have been observed for electrons flowing in spin-orbit coupled materials such as GaAs or InGaAs\cite{Awschalom2004, Jungwirth2005} and for laser light traversing dielectric junctions\cite{Kwiat2008}.  Here we observe the spin Hall effect in a quantum-degenerate Bose gas, and use the resulting spin-dependent Lorentz forces to realize a cold-atom spin transistor.  By engineering a spatially inhomogeneous spin-orbit coupling field for our quantum gas, we explicitly introduce and measure the requisite spin-dependent Lorentz forces, in excellent agreement with our calculations.  This atomtronic  circuit element behaves as a new type of velocity-insensitive adiabatic spin-selector, with potential application in devices such as magnetic\cite{Donley2011} or inertial\cite{Galitski2011} sensors.  In addition, such techniques --- for both creating and measuring the SHE --- are clear prerequisites for engineering topological insulators\cite{Mele2005,Zhang2006} and detecting their associated quantized spin Hall effects in quantum gases. As implemented, our system realized a laser-actuated analog to the Datta-Das spin transistor\cite{Juzeliunas2008, Das1990}.  

\end{abstract}

The spin Hall effect (SHE) is generated by spin-dependent forces transverse to a particle's motion --- akin to the Lorentz force --- that can act on electrons\cite{Awschalom2004, Jungwirth2005, olejnik2012}, photons\cite{Kwiat2008}, or as here, neutral atoms.  Each of these has an internal, or ``spin,'' degree of freedom that can be either up or down, creating a (pseudo-)spin 1/2 system.  In materials, microscopic spin-orbit coupling (SOC) induces the SHE in one of two primary ways: via an intrinsic mechanism driven directly by the SOC\cite{MacDonald2004}, or via an extrinsic mechanism which additionally requires scattering from impurities\cite{Perel1971, Hirsch1999}. The motion of spins in systems with a SHE is strikingly similar to the motion of charges in an external magnetic field, but with equal and opposite effective Lorentz forces for each of the two spin states. Thus, just as the Lorentz force gives rise to the Hall effect for charged particles, spin-dependent Lorentz forces (SDLFs) generate spin Hall effects. 


In the Hamiltonian description of quantum mechanics, forces are described in terms of associated potentials. For example, a magnetic field $\bm{B} = \bm{\nabla}\times\bm{A}$ is generated from a vector potential $\bm{A}$ that enters into the Hamiltonian $\bm{\hat{H}} = (\bm{\hat{p}} -q_0\hat{\bm{A}})^2/2m$ with canonical momentum $\bm{\hat{p}}$, charge $q_0$, and mass $m$. We engineered a vector potential $\bm{A}$ that depends on an effective spin degree of freedom with opposite sign for the two effective spin states. This can create a SDLF and SHE when the spins move perpendicular to the resulting spin-dependent $\bm{B}$.

More formally, this vector potential can be expressed as a vector of $2 \times 2$ matrices, leading to a generalized relationship between the vector potential $\bm{\mathcal{\check{A}}}$ and magnetic field\cite{Fleischhauer2005}
\begin{equation}
\label{eq:bfield}
\bm{\mathcal{\check{B}}} = \bm{\nabla}\times\bm{\mathcal{\check{A}}}- \frac{i}{\hbar}\bm{\mathcal{\check{A}}}\times\bm{\mathcal{\check{A}}}.
\end{equation}
The Heisenberg equations of motion show that $\bm{\mathcal{\check{B}}}$ is the generalized magnetic field in a spin-dependent Lorentz force law (Methods). The first term in eq.~\ref{eq:bfield} is analogous to the conventional magnetic field, while the second term is only non-zero when the vector components of $\bm{\mathcal{\check{A}}}$ do not all commute, i.e., $\bm{\mathcal{\check{A}}}$ is non-Abelian. The generalized Lorentz force for the two spin states can be equal and opposite, for example, when $\bm{\mathcal{\check{B}}} = \mathcal{B}_0 \check{\sigma}_3 {\bf e}_z$, where $\mathcal{B}_0$ describes the field's magnitude and $\check{\sigma}_{1, 2, 3}$ are the $2 \times 2$ Pauli matrices.

There are two different classes of vector potentials (unrelated by gauge transformations) that lead to this magnetic field, each exploiting different terms in eq.~\ref{eq:bfield}. For example, in 2D material systems, almost every possible form of linear SOC --- combinations of linear Dresselhaus\cite{Dresselhaus1955} or Rashba\cite{Rashba1984} --- is equivalent to a spatially uniform non-Abelian vector potential with $ - i \left(\bm{\mathcal{\check{A}}} \times \bm{\mathcal{\check{A}}}\right) / \hbar \propto \check{\sigma}_3 {\bf e}_z$ (see Methods and Ref.~\onlinecite{ LyandaGeller1998}). In contrast, we engineered a spin-orbit coupled Hamiltonian with a spatially-dependent Abelian vector potential that produces $\bm{\nabla}\times\bm{\mathcal{\check{A}}}  \propto \check{\sigma}_3 {\bf e}_z$.


The relationship between these two distinct vector potentials  is unusual. While the equations of motion describe the same SDLF leading to an intrinsic SHE, the associated energy spectra are different (e.g. in the 2D material systems discussed above, $\left[ \bm{\mathcal{\check{B}}}, \bm{\hat{H}}  \right] \neq 0$, implying that $\bm{\mathcal{\check{B}}}$ is time-dependent in the Heisenberg picture). Still, both can give rise to time-reversal (TR) invariant topological insulators (TIs). The $ - i \left(\bm{\mathcal{\check{A}}} \times \bm{\mathcal{\check{A}}}\right) / \hbar$ case mirrors the typical situation in materials where the intrinsic SOC leads to topological band structure\cite{Zhang2006}. The $\bm{\nabla}\times\bm{\mathcal{\check{A}}}$ case leads to the most simple conceptual example of a TI: two superimposed quantum Hall systems with equal but opposite magnetic fields\cite{Kane2010} (a single quantum Hall system is a TI, but with broken TR symmetry). Both types of vector potentials exhibit the quantum spin Hall effect leading to TIs; the latter type of TI is impractical in material systems, but is a direct extension of the quantum gas SHE demonstrated in this work\cite{DasSarma2009} (Methods and Supplementary Information).

\begin{figure}[t!]
\begin{center}
\includegraphics[scale=1]{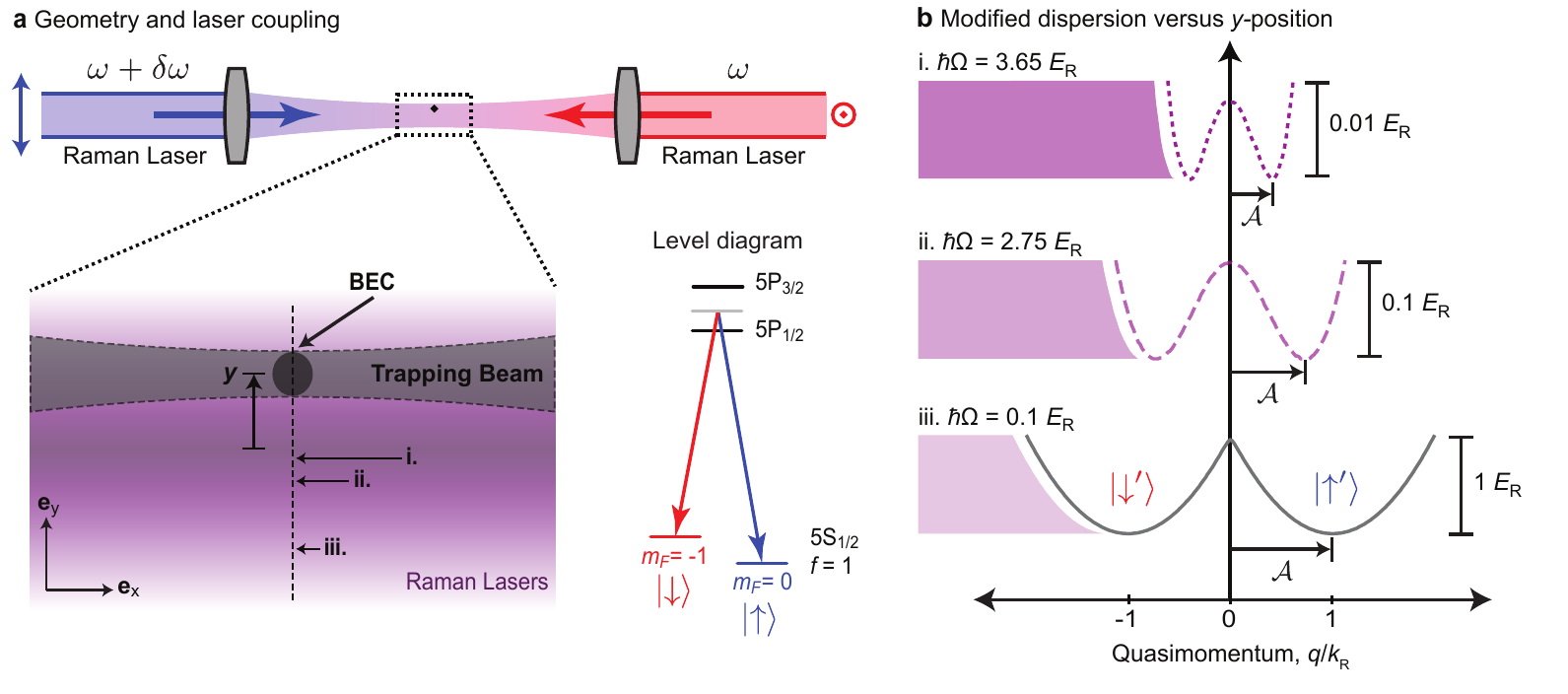}
\end{center}
\caption{{\bf Experiment schematic.} {\bf a} Raman beams with frequencies $\omega$ and $\omega+\delta\omega$ propagating along \ex~coupled two states in $^{87}$Rb's $f=1$ ground state manifold. Dynamic control of an optical trapping beam propagating along \ex~allowed for movement of the BEC along \ey, giving a time- and position-dependent Raman coupling. The Raman coupling altered the free-particle dispersion along \ex, creating double wells\cite{Spielman2011} in quasimomentum $q$. These modified dispersions $E(q)$ are shown in {\bf b} for the three different $y$-positions marked in {\bf a}. We associate states near the minimum of each well with dressed spins, and identify the location of the minima with a synthetic vector potential $\mathcal{A}$.
\label{fig:setup}}
\end{figure}

We realized the SHE with ultracold atoms following the proposal of Ref.~\onlinecite{Duan2006} by subjecting pseudospin-1/2 $^{87}$Rb Bose-Einstein condensates (BECs) to spin- and space-dependent vector potentials. Two ``Raman lasers'' with wavelength $\lambda$, counterpropagating along \ex, coupled the $\left|f =1; m_F = 0, -1\right> = \ket{ \uparrow, \downarrow}$ spin states comprising our pseudo-spin-$1/2$ system (in analogy to the spin-$1/2$ electron) with strength $\Omega$ (Fig.~\ref{fig:setup}a). $\lambda$ determines the single-photon recoil energy $E_{\rm{R}} = \hbar^2 k_{\rm{R}}^2 / 2 m$, momentum $\hbar k_{\rm{R}} = 2 \pi \hbar / \lambda$ and velocity $v_{\rm{R}} = \hbar  k_{\rm{R}} /m$, where $m$ is the mass of a $^{87}$Rb atom and $2 \pi \hbar$ is Planck's constant. In this configuration, the Hamiltonian describing motion along \ex~includes an effective SOC term\cite{Spielman2011, Pan2012, Zhang2012, Zwierlein2012}, altering the dispersion relation as shown in Fig.~\ref{fig:setup}b. This modified dispersion features two degenerate wells each displaced from zero by an amount $\mathcal{A} =  k_{\rm{R}} \left[ 1 - \left(\hbar \Omega/ 4 E_{\rm{R}} \right)^2\right]^{1/2}$ for $\hbar \Omega < 4 E_{\rm{R}}$ (see Methods Summary). Particles with momenta near these minima can be thought of as dressed spin states $ \ket{ \uparrow^\prime, \downarrow^\prime }$ (which we will colloquially refer to as spin states) in the presence of a vector potential  $\bm{\mathcal{\check{A}}} = \mathcal{A} \check{\sigma}_3 {\bf e}_x$. Given that $\Omega$ depends on the intensity of the Raman lasers, $\mathcal{A}$ inherits the spatial dependence of the Raman lasers' Gaussian intensity profile. 

\begin{figure}[t!]
\begin{center}
\includegraphics[scale=1]{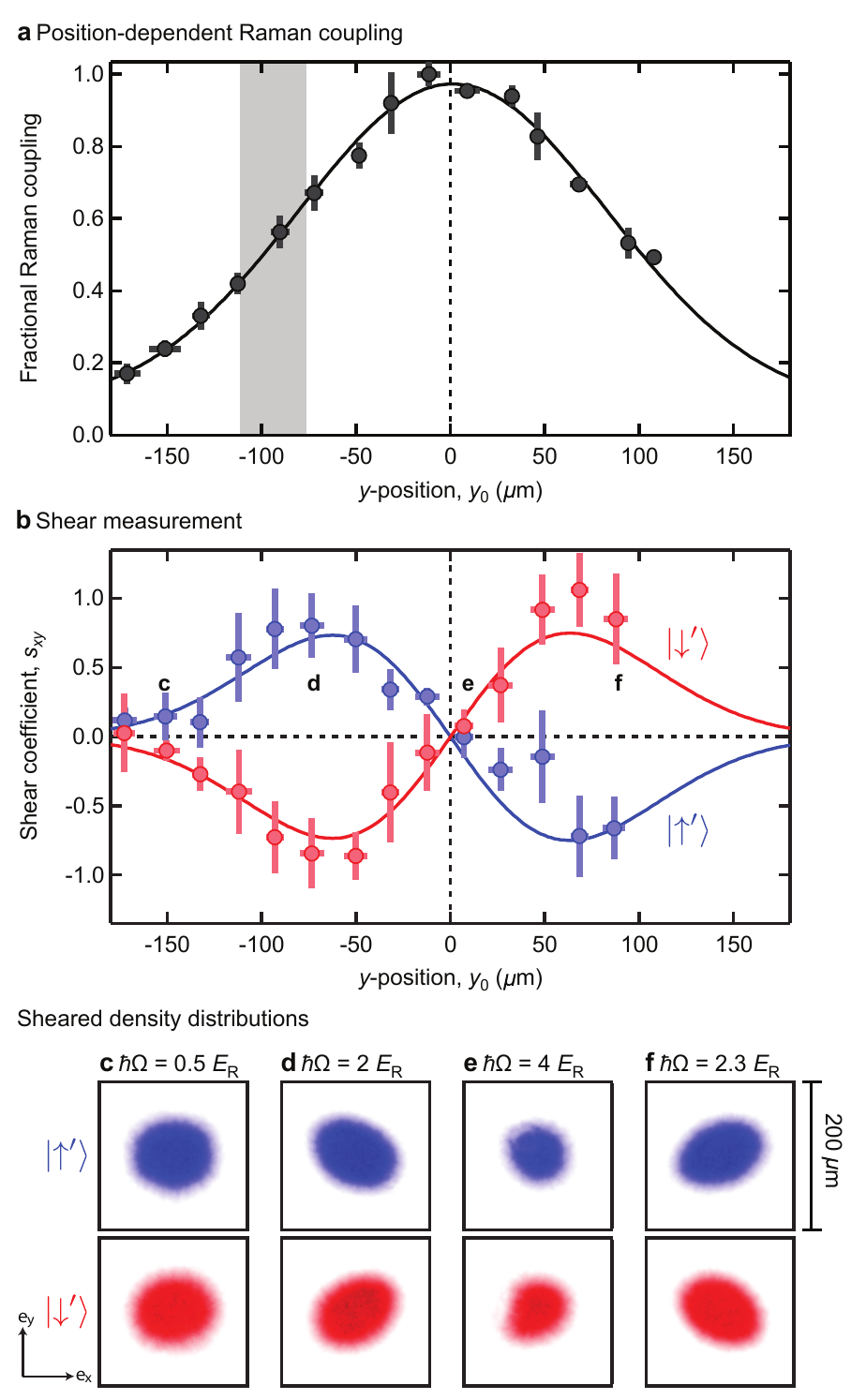}
\end{center}
\caption{
{\bf Spin Hall currents.} {\bf a} Raman coupling strength versus $y$-position, fit with the Raman laser's Gaussian profile. In this figure, uncertainties reflect the standard deviation of $\approx 5$ measurements. {\bf b} The observed shear coefficient $s_{xy}$ (see text) was opposite for each spin and the magnitude followed the derivative of a Gaussian function (solid lines). {\bf c-f} Representative 2D spin-momentum distributions observed after TOF at different $y$-positions. For the data in {\bf b - f}, the magnitude of the effect was enhanced by elongating the BEC along \ey~(Methods), sampling a greater range of the vector potential. The measurements in Figs.~\ref{fig:throw} and \ref{fig:twospins} were taken in the portion of the laser shaded in gray in {\bf a}, where the curl of $\mathcal{A}$ is large and nearly uniform.
}\label{fig:skew} 
\end{figure}

\begin{figure}[b!]
\begin{center}
\includegraphics[scale=1]{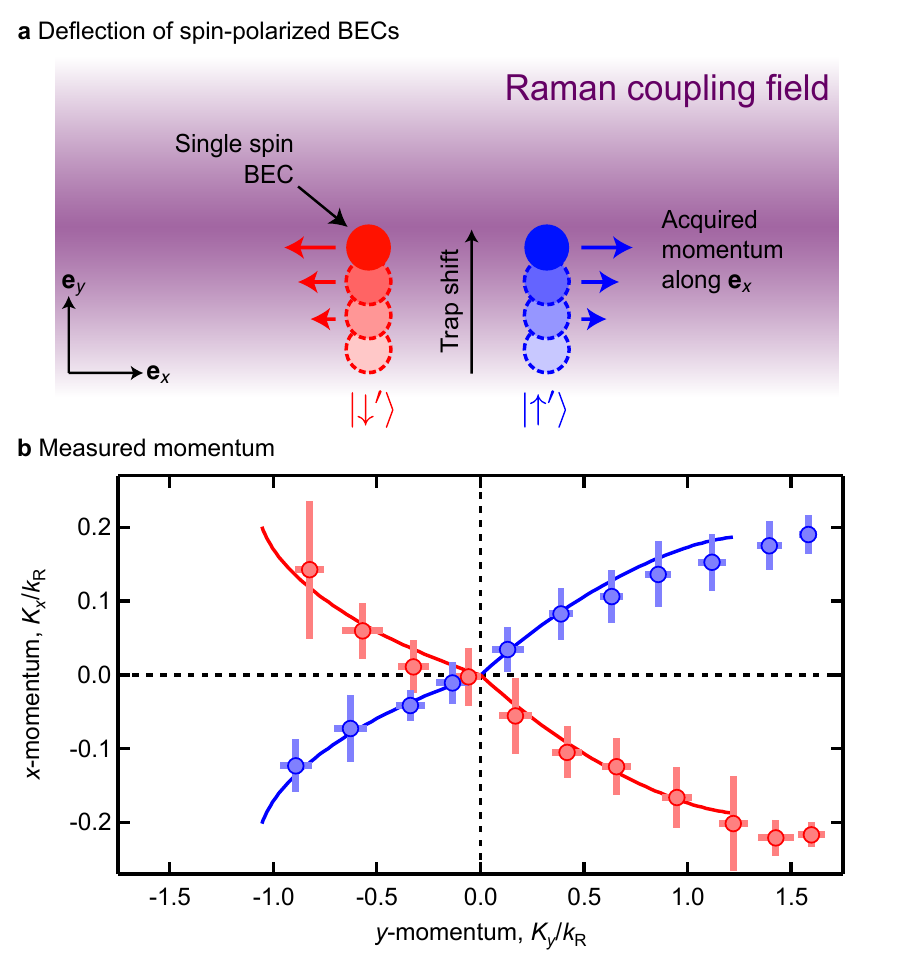}
\end{center}
\caption{
{\bf Spin-polarized SHE.} {\bf a} Spin-dependent forces along \ex~from motion along \ey.  {\bf b} Acquired momentum along \ex~versus final momentum along \ey: blue circles denote $\ket{\uparrow^\prime}$ and red circles denote $\ket{\downarrow^\prime}$.  The solid curves are solutions of Heisenberg's equations of motion for each spin, fit to the data with $\Omega$ as the only free parameter (Methods). The resulting $\Omega$ is within 15$\%$ of our measured coupling strength of 2.5(2) $E_{\rm{R}}$ at $y = -115 \micron$, the center of the spatial region sampled by the atoms during the measurements (gray shaded region in Fig.~\ref{fig:skew}a).  Uncertainties reflect the standard deviation of $\approx 5$ measurements.
}\label{fig:throw}
\end{figure}

\begin{figure}
\begin{center}
\includegraphics[scale=1]{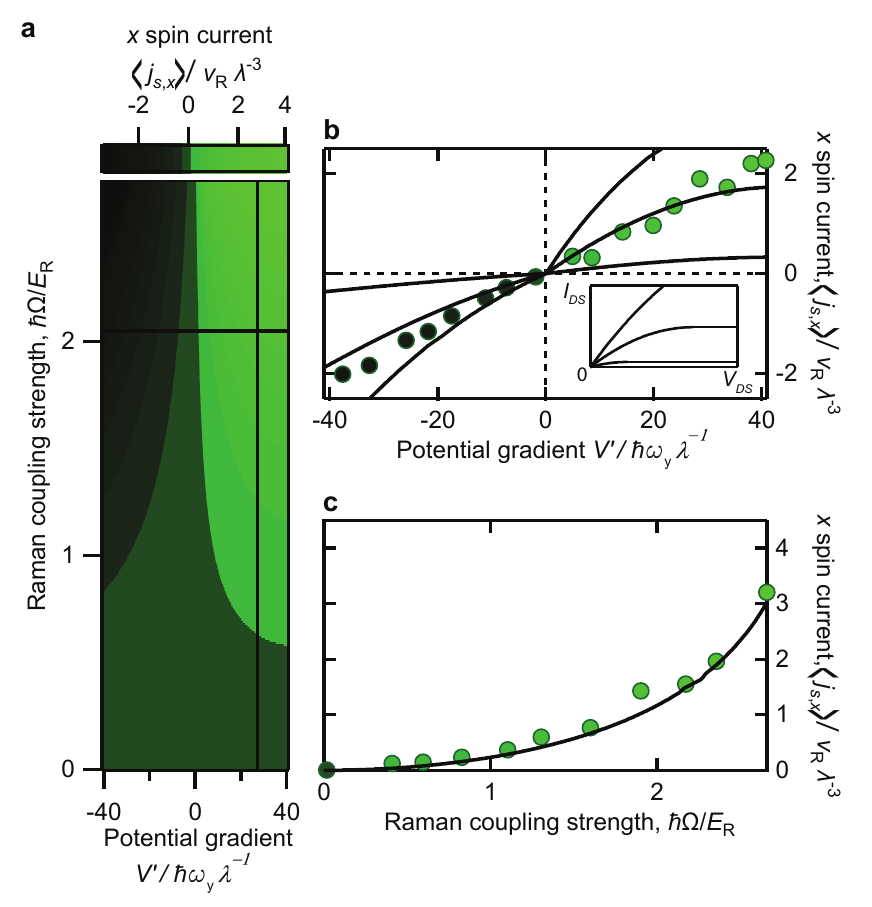}
\end{center}
\caption{
{\bf Spin Hall currents.} {\bf a} Calculated spin current  versus  potential gradient $V'$ and coupling strength $\Omega$. The two cuts (black lines) show the parameters at which measurements of $\left<j_{s,x}\right>$ were made. {\bf b} Spin current $\left<j_{s,x}\right>$ versus $V'$. Note, $\hbar\Omega(y = -115~\micron) = 2.3(2)~E_{\rm{R}}$. The central solid curve is a fit of our model to the data, with fit value $\hbar \Omega_f = 2.05~E_{\rm{R}}$. The remaining curves are the modeled response at $\hbar \Omega = 1~E_{\rm{R}}$ (lower magnitude) and 2.5 $E_{\rm{R}}$ (higher magnitude). Inset:  FET drain-source current $I_{DS}$ versus drain source voltage $V_{DS}$ for three $V_{GS} = $1.5, 2.4, and 3.5 V above threshold.  {\bf c} Spin current dependence on coupling strength at a fixed trap displacement from $y_0 = -122 \micron$  to $y_f = -95 \micron$ with $V' \approx 27~\lambda /\hbar \omega_y$. The solid curve is the fit of our model to the data (Methods). The scatter in the data is reflective of typical uncertainties. 
}\label{fig:twospins}
\end{figure}

The spatial dependence of $\mathcal{A}$ gives rise to a SHE in our quantum gas\cite{Duan2006, Oh2007}. To probe the mechanism underlying the SHE, we abruptly changed $\mathcal{A}$ and observed spin-dependent shearing of the atomic cloud (Fig.~\ref{fig:skew}). We then observed --- for a time-independent $\mathcal{A}$ --- the resulting SHE using two techniques: (i) we propelled atoms in either $ \ket{ \uparrow^\prime }$ or  $ \ket{ \downarrow^\prime }$ along \ey~and detected a spin-dependent Lorentz-like response along $\pm$\ex~(Fig.~\ref{fig:throw}); and (ii) using a mixture of both dressed spins, we used the SDLF to realize a spin transistor~(Fig.~\ref{fig:twospins}).

These experiments began with $5 \times 10^4$ atom BECs prepared in $\ket{\uparrow}$, $\ket{\downarrow}$, or mixtures thereof, confined in a crossed-beam optical dipole trap with typical frequencies $(\omega_x, \omega_y, \omega_z)/2\pi$ $\approx (35, 35, 100) \Hz$. The $\lambda = 790.13\nm$ Raman laser beams, traveling along $\pm{\bf e}_x$, had $170\micron$ waists ($1/e^2$ radius).  We moved the BECs along \ey, sampling this inhomogeneous Raman laser profile, by displacing the appropriate trap beam.  At any given initial $y$-position $y_0$, we then adiabatically turned on the Raman lasers in $150\ms$, Raman-dressing the BEC\cite{Spielman2011} and transforming our initial spin states into their dressed counterparts, at rest\cite{CZhang2012, Zhang2011} (Methods). 

We explored the spin- and space-dependence of the vector potential $\bm{\mathcal{A}}(y)$ by observing the response of BECs to abrupt temporal changes in $\mathcal{A}$. When $\mathcal{A}$ depended on $y$, these changes sheared the BECs' density distribution. We prepared spin-polarized BECs at a variable position $y_0$ (Methods).   Each Raman-dressed BEC therefore sampled a range of Raman coupling strengths across its $40\micron$ diameter (Fig.~\ref{fig:skew}a).  Upon suddenly turning off the Raman lasers, the BEC --- initially at rest --- experienced a spin-dependent ``electric'' force $-\partial \bm{ \mathcal{A}}/\partial t$ resulting from a time-changing vector potential along \ex\cite{Spielman2011b}.  We probed this system by switching off the dipole trap and the Raman beams in less than $1\us$ and absorption-imaging the atoms after a $30\ms$ time-of-flight (TOF, common to all of our measurements).  


As $\bm{\mathcal{A}}(y)$ depended on both spin and $y$-position, we observed a spin- and $y_0$-dependent shear\cite{Spielman2012} in the density distribution (Fig.~\ref{fig:skew}b) after TOF, described by $n(x,y,z) \propto 1- (x/R_x) ^2 - (y/R_y)^2 - (z/R_z)^2 - s_{xy} x y / (R_x R_y)$, where $R_{x,y,z}$ are the Thomas-Fermi radii. The spatial dependence of $\mathcal A$ is quantified by the shear coefficient $s_{xy}$ obtained by fitting this distribution (integrated along \ez) to the TOF BEC density distribution. The spin-dependent nature of the vector potential is evident in the opposite sign of the shear for each spin (Fig. \ref{fig:skew}b-f) and in that the magnitude of the shear coefficient follows the local derivative of the vector potential at the BEC's center.

We first observed the SHE using spin-polarized BECs. This would be atypical in condensed matter systems, where both spins are usually present. After preparing a spin-polarized BEC at a position $y_0$ between $y_{\rm{min}} = -135~\micron$ and $y_{\rm{max}} =  -95~\micron$ (gray shaded region in Fig.~\ref{fig:skew}a, a region over which the SDLF was both reasonably large and uniform), we suddenly displaced the center of the harmonic trap to $y_{f}$, either $y_{\rm{max}}$ or $y_{\rm{min}}$. This displacement can formally be understood as resulting from an applied potential with gradient $V'$. The atoms accelerated to a final $y$-momentum $\hbar K_y$ in $\approx 7\ms$ (1/4 of the \ey-trap period). During this time, the SDLF accelerated the atoms perpendicular to their instantaneous momentum, resulting in a final $x$-momentum $\hbar K_x$. By waiting this quarter-period after trap displacement, we ensured that the atoms always arrived at $y_f$ (regardless of $y_0$). Subsequently, the trap was turned off suddenly ($t_{\rm{off}} <1 \us$), the Raman lasers were turned off slowly compared to dressed state band gaps ($\sim$500 $\us$), and the atoms were imaged after TOF to determine their final momentum (Fig.~\ref{fig:throw}b). With this turn-off procedure, the atoms experienced a force $ -\partial \bm {\mathcal{A}}/\partial t$ along \ex~(independent of $y_0$) that offset the final center of mass position after TOF (Methods). We calibrated this zero-momentum TOF-position by detecting atoms released from rest at $y_f$.


Each spin-polarized BEC acquired a momentum along \ex~that was directed oppositely for the two spins and related to its final momentum along \ey, demonstrating an intrinsic spin Hall effect. We modeled the dynamics of each spin (Methods, solid curves in Fig.~\ref{fig:throw}b) by solving the Heisenberg equations of motion. As our atoms remain in the lowest-energy band plotted in Fig.~\ref{fig:setup}b, the Heisenberg equations of motion reduce to classical dynamics subject to Fig.~\ref{fig:setup}b's spin-orbit-coupled dispersion curves. The model predicts both $K_x$ and $K_y$ as a function of initial and final trap displacement. We leave  $\Omega$ as a fit parameter, the value of which is within $15\%$ of our calibrated value. The results of this model are plotted along with the data in Fig. \ref{fig:throw}b.


Next, we realized the spin Hall effect in a configuration analogous to solid systems by using mixtures of both spins. In the presence of both spins, we define average spin and particle current densities $\left<\textbf{j}_s\right> = \left<\textbf{j}_{\uparrow^\prime}\right>-\left<\textbf{j}_{\downarrow^\prime}\right>$ and  $\left<\textbf{j}_p\right> = \left<\textbf{j}_{\uparrow^\prime}\right>+\left<\textbf{j}_{\downarrow^\prime}\right>$, where the average current density for spin $i$ (either $\uparrow^\prime$ or $\downarrow^\prime$) is $\left<\textbf{j}_{i}\right>= \int_V{n_i(\textbf{r})\textbf{v}_i(\textbf{r})\rm{d}\textbf{r}}  / V$, with density $n$, velocity $\textbf{v}$, and in-situ BEC volume $V$. An equal current of each spin moving in the same direction corresponds to a pure particle current, while an equal current of each spin moving in opposite directions gives a pure spin current.

This third class of experiments started with BECs in a mixture of both spins (Methods). We generated a pure particle current using the trap displacement technique described above. As before, the system evolved under the SDLF for $\approx$ 7 ms, after which time the atoms were released from the trap and the Raman lasers adiabatically turned off (Methods). Each TOF image contained information about both dressed spin states, allowing us to simultaneously determine the spin and particle currents. We modeled the resulting spin current along \ex~as a function of coupling strength $\Omega$ and potential gradient $V'$, and Fig. \ref{fig:twospins}a shows the system's spin response $\left<j_{s,x}\right> = \left<{\bf j}_s\right> \cdot {\bf e}_x$. By varying one parameter at a time (black lines in Fig. \ref{fig:twospins}a), we measured the spin current as a function of $V'$ (Fig.~\ref{fig:twospins}b) or as a function of $\Omega$ (Fig. \ref{fig:twospins}c). In both cases the experiment agrees with our model.


Despite the universal existence of the SHE in spin-orbit coupled metals and semiconductors, the technology for studying the spin Hall effect was developed only recently. Soon afterwards, the SHE was exploited to develop spintronic devices\cite{olejnik2012}. In this spirit, our experiment describes an externally actuated ``atomtronic'' bipolar spin transistor\cite{Juzeliunas2008, Das1990}, where $\Omega$ plays the role of the transistor's gate voltage and the potential gradient $V'$ is analogous to the drain-source voltage. The spin current turns on sharply at $\hbar \Omega \approx 1~E_{\rm{R}} $ (Fig. \ref{fig:twospins}c), with a final spin current set by the potential gradient. Meanwhile, for a given Raman coupling (``gate voltage''), the spin current turns on smoothly with positive or negative particle current (Fig. \ref{fig:twospins}b). This similarity between our system and a field-effect transistor (FET) is further highlighted in Fig.~\ref{fig:twospins}b, where the three black curves modeling our system's response at three different Raman coupling strengths are compared to the characteristic response of an FET's drain current $I_D$ as a function of drain-source voltage $V_{DS}$ at three different gate-source voltages $V_{GS}$.

In atomic systems, other techniques can separate particles by spin, such as the well-known Stern-Gerlach (SG) effect. Our technique complements these, as the spin-dependent force depends not on the atoms' position (as in the SG effect), but on their velocity. For example, a SHE device with a finite interaction region will deflect an incoming atomic beam by an amount independent of its entering velocity; though an increase in initial velocity  decreases the interaction time, the perpendicular force increases (for interaction times much less than $2 \pi m / \bm{\mathcal{\hat{B}}}$). For devices using the SG effect, the deflection depends only on the interaction time, which changes with initial velocity. A spin transistor might operate using either our SDLF or a SG-type force, but its behavior will be quite different.  For example, using our transistor as the input and output beam splitter in Mach-Zehnder-type inertial sensors\cite{Donley2011} could yield coherent adiabatic momentum splitting that is independent of the atoms' longitudinal velocity profile.



We demonstrated an intrinsic spin Hall effect in a quantum gas using a precisely engineered spin- and space- dependent vector potential. Systems such as this --- with the currently available experimental parameters --- are candidates for ac gravity gradiometers, when applied to dilute clouds where interaction effects are negligible\cite{Galitski2011}. In addition, TR invariant TIs manifest the quantum spin Hall effect (QSHE)\cite{Kane2010}. Using current technologies, our method for producing the SHE could produce the QSHE in an ultracold gas of fermionic $^{40}$K (Methods, Supplementary Material). Despite the technical challenges, the simplicity of our setup --- two atomic spin states and two counterpropagating lasers --- makes our approach an appealing method for achieving the QSHE. In similar parameter regimes, a Bose gas may realize exotic interacting topological insulators\cite{DasSarma2009, Ohberg2011}.



\begin{methodssummary}
\subsection{System Preparation}
A $B_0 = 2.1\mT$ bias magnetic field lifted the degeneracy of the $\ket{f=1, m_F = 0, \pm 1}$ spin states in $\Rb87$'s electronic ground state manifold, leading to an energy level splitting $\Delta E = 2 \pi \hbar \times 15~\rm{J}$  between $\ket{m_F= -1}$ and $\ket{m_F= 0}$,  matching the $\hbar\, \delta \omega$ energy difference between the Raman laser beams' photons. Due to the large bias field, the  $\ket{m_F= +1}$ spin state was detuned from Raman resonance by 17.8 $E_{\rm{R}}$, and was inactive in our experiments.

In the limit of zero Raman coupling, each dressed spin continuously connects to a bare spin with quasimomentum $|q| =~1~k_R$. To load a specific dressed spin, we started with a BEC in $|m_F= -1\mathinner\rangle$, $|m_F= 0\mathinner\rangle$,  or a mixture thereof, and turned on the Raman lasers in $150$ ms. During spin-polarized experiments, we prevented the undesired population of the other dressed spin by applying a detuning $\hbar \delta = \Delta E - \hbar \, \delta\omega =  0.15 E_{\rm{R}}$ during the ramp up of $\Omega$, then shifting to resonance ($\delta = 0$) with a 1 ms ramp of $B_0$. An acousto-optic modulator shifted the position of the dipole trap beam propagating along \ex, allowing controlled translation of the atomic sample along \ey.

\subsection{Dressed states}
The single particle properties of our system are well-described by the Hamiltonian\cite{Spielman2011}
\begin{equation}
\hat{H} = \frac{\hbar^2 \left(  \hat{q}^2+\hat{k}_y^2+\hat{k}_z^2  \right) }{2 m}\check{\mathbb{1}}+\frac{\hbar \Omega}{2}\check{\sigma}_1 -  \frac{\hbar^2 k_{\rm{R}} \hat{q}}{m}\check{\sigma}_3 + E_{\rm{R}} \check{\mathbb{1}}
\end{equation}
for resonant Raman coupling, as we use here. The eigenenergies 
\begin{equation}
\mathcal{E}_{\pm}(q) +\frac{\hbar^2 \left( k_y^2+k_z^2  \right) }{2 m}, \rm{with}~ \mathcal{E}_{\pm}(q) = \frac{\hbar^2 q^2}{2 m}+E_{\rm{R}} \pm \sqrt{\left(\frac{\hbar \Omega}{2}\right)^2+\left(  \frac{\hbar^2 k_{\rm{R}} q }{m}  \right)^2}
\end{equation}
define a pair of effective dispersion relations, the lower of which, $\mathcal{E}_-(q)$, is plotted for $k_y=k_z = 0$ in Fig. \ref{fig:setup}b for a selection of coupling strengths.
\end{methodssummary}



\bibliography{bibliography}


\begin{addendum}
\item This work was partially supported by the DARPA OLE program; the ARO's atomtronics MURI, NIST, and the NSF through the PFC at the JQI. M.C.B. acknowledges NIST - ARRA; L.J.L. acknowledges support from NSERC; K.J.-G. acknowledges CONACYT.  
\item[Author Contributions] M.C.B. led the data taking effort in which all co-authors participated. M.C.B. carried out the analysis, M.C.B. and I.B.S. performed theoretical and analytical calculations, and all authors contributed to writing the manuscript.

\item[Competing Interests]  The authors declare that they have no competing financial interests.

\item[Correspondence] Correspondence and requests for materials should be addressed to I.B.S.~(email: ian.spielman@nist.gov).

\end{addendum}


\begin{methods}

\subsection{System Preparation}
A $B_0 = 2.1\mT$ bias magnetic field lifted the degeneracy of the $\ket{f=1, m_F = 0, \pm 1}$ spin states in $\Rb87$'s electronic ground state manifold, leading to an energy level splitting $\Delta E = 2 \pi \hbar \times 15~\rm{J}$  between $\ket{m_F= -1}$ and $\ket{m_F= 0}$,  matching the $\hbar\, \delta \omega$ energy difference between the Raman laser beams' photons. Due to the large bias field, the  $\ket{m_F= +1}$ spin state was detuned from Raman resonance by 17.8 $E_{\rm{R}}$, and was inactive in our experiments.

In the limit of zero Raman coupling, each dressed spin continuously connects to a bare spin with quasimomentum $|q| =~1~k_R$. To load a specific dressed spin, we started with a BEC in $|m_F= -1\mathinner\rangle$, $|m_F= 0\mathinner\rangle$,  or a mixture thereof, and turned on the Raman lasers in $150$ ms. During spin-polarized experiments, we prevented the undesired population of the other dressed spin by applying a detuning $\hbar \delta = \Delta E - \hbar \, \delta\omega =  0.15 E_{\rm{R}}$ during the ramp up of $\Omega$, then shifting to resonance ($\delta = 0$) with a 1 ms ramp of $B_0$. An acousto-optic modulator shifted the position of the dipole trap beam propagating along \ex, allowing controlled translation of the atomic sample along \ey.

\subsection{Dressed states}
The single particle properties of our system are well-described by the Hamiltonian\cite{Spielman2011}
\begin{equation}
\hat{H} = \frac{\hbar^2 \left(  \hat{q}^2+\hat{k}_y^2+\hat{k}_z^2  \right) }{2 m}\check{\mathbb{1}}+\frac{\hbar \Omega}{2}\check{\sigma}_1 -  \frac{\hbar^2 k_{\rm{R}} \hat{q}}{m}\check{\sigma}_3 + E_{\rm{R}} \check{\mathbb{1}}
\end{equation}
for resonant Raman coupling, as we use here. The eigenenergies 
\begin{equation}
\mathcal{E}_{\pm}(q) +\frac{\hbar^2 \left( k_y^2+k_z^2  \right) }{2 m}, \rm{with}~ \mathcal{E}_{\pm}(q) = \frac{\hbar^2 q^2}{2 m}+E_{\rm{R}} \pm \sqrt{\left(\frac{\hbar \Omega}{2}\right)^2+\left(  \frac{\hbar^2 k_{\rm{R}} q }{m}  \right)^2}
\end{equation}
define a pair of effective dispersion relations, the lower of which, $\mathcal{E}_-(q)$, is plotted for $k_y=k_z = 0$ in Fig.~1b for a selection of coupling strengths.

\subsection{Quantum Spin Hall Effect}
Our technique for producing the spin Hall effect can be extended to realize the quantum spin Hall effect (QSHE) in 2D ultracold Fermi gases. A simple example system that exhibits the quantum spin Hall effect can be constructed by overlapping two filling factor $\nu=1$ integer quantum Hall (IQHE) systems with opposite magnetic field and therefore opposite Chern numbers\cite{Kane2010}. While this construct --- spatially overlapping two separate electron systems that each experience an opposite magnetic field --- is artificial, the quantum spin Hall effect can arise from an equal mixture of spins experiencing strong opposite spin-dependent ``magnetic'' fields.

To understand how this might work intuitively, consider our effective pseudospin Hamiltonian in 2D for $\hbar \Omega < 4~E_{\rm{R}}$ (ignoring the optical confinement, the scalar light shift from the Raman lasers, and the zero-energy shift from the Raman dressing)
\begin{equation*}
\label{pseudoham}
\hat{H} = \frac{1}{2 m^*}\left(\bm{\hat{p}}\check{\mathbb{1}}- \mathcal{A} \check{\sigma}_3 \bm{e}_x      \right)^2,
\end{equation*}
with $\check{\mathbb{1}}$ the $2 \times 2$ identity matrix, $\mathcal{A}=\hbar k_{\rm{R}} \left[   1-\left(  \hbar \Omega  /  4 E_{\rm{R}}\right)^2  \right]^{1/2}$ the Raman laser-induced vector potential, $\bm{\hat{p}}$ the canonical momentum, and $m^*$ the effective mass tensor of our particles. Here, pseudospin is a good quantum number and the system can be thought of as two independent systems that respond oppositely to temporal and spatial gradients of $\mathcal{A}$. By introducing a large non-zero curl for $\mathcal{A}$, each spin state taken separately could be driven to the IQHE regime, thereby creating a QSHE in a system composed of an equal mixture of both spins. 

Our specific proposal to extend our work and realize the QSHE uses $^{40}$K confined in a quasi-2D geometry in the $\bm{e}_x - \bm{e}_y$ plane. Two Raman lasers counter-propagating along $\bm{e}_x$ couple together two magnetic sublevels in the $\ket{f=9/2}$ ground state manifold. Tailoring the Raman lasers (using a spatial light modulator\cite{Jesper2007, DeMarco2008a, Hadzibabic2012a, Hill2013 }, for instance) to have a position-dependent coupling $\hbar \Omega(y) = 4 E_{\rm{R}} \sqrt{L_y^2-y^2}/ L_y$ for $y \in (0, L_y]$ along $\bm{e}_y$ produces a linearly-varying $\mathcal{A}$. Each pseudospin experiences an oppositely-directed uniform synthetic magnetic field with cyclotron frequency $\omega_c = \hbar k_{\rm{R}}/m L_y$ for $y \in (0, L_y]$. 

To reach the QSHE regime, the thermal energy scale $k_{\rm{B}} T$, Fermi energy $\mathcal{E}_{\rm{F}}$, and cyclotron energy $\hbar \omega_c$ must satisfy $k_{\rm{B}} T < \mathcal{E}_{\rm{F}} \approx \hbar \omega_c$ (so that the Fermi energy falls in the gap between the ground and first Landau-levels). Here, $k_{\rm{B}}$ is Boltzmann's constant and $T$ is the temperature. The cyclotron frequency therefore sets the energy scales necessary to see a QSHE. For realistic system sizes of 5 -- 10 $\mu$m, the cyclotron frequency is  $\omega_c / 2 \pi \approx 100$~Hz.  In the Online Supplementary Materials we make this argument rigorous for our actual experimental configuration.

\subsection{Figure 2 Notes}

The Raman coupling strength in Fig. 2a was measured as described in refs.~\onlinecite{Zhang2012, Spielman2009b}. For the data in Figs. 2b-f, the aspect ratio of the BEC was adjusted from its typical cylindrical symmetry to be $50 \%$ longer along \ey~than \ex~by adjusting the optical trap, and the atom number was maintained $> 10^5$.

\subsection{Measurement and Analysis}

To measure the atoms' momenta, the optical confinement was turned off suddenly while the Raman lasers' intensity was linearly ramped to zero in 0.5 to 1 ms. This procedure transferred each dressed spin to a bare spin moving with an $x$-momentum equal to its quasimomentum $\hat{q}$ and a $y$-momentum equal to its in-trap $y$-momentum $K_y$. A magnetic field gradient applied for a few ms during the 30 ms TOF separated the two bare spins along \ey~via the Stern-Gerlach effect, after which we measured the atomic density distribution and obtained its mean position. To determine the atoms' in situ momenta, we referenced the measured TOF positions to the TOF positions observed for atoms under the same experimental conditions, but at rest. For example, when the trap was suddenly displaced as in Figs.~3 or 4, the reference position was determined by adiabatically dressing the atoms at the final trap position and measuring the TOF position. Subtracting the TOF position of the suddenly-displaced atoms from the reference TOF position allowed us to determine the in-trap momentum.

This measurement of the momenta contained two contributions which biased the TOF positions away from the actual momentum. If the atoms do not reach their equilibrium position in the trap before TOF begins, our subtraction procedure does not yield the actual velocity, as this initial displacement is interpreted as momentum after TOF. According to our simulations, this resulted in a systematic underestimation of the momentum along \ex~and \ey. In addition, to compensate gravity during displacement of the optical trap, the overall intensity of the optical trapping beams was increased by 25$\%$ at the same time the position of the optical trap was changed. Due to the competition between the optical trap and the near-linear spatial dependence of the energy minimum of the Raman-dressed bands, this power increase shifted the equilibrium position of the atoms along \ey~even in the absence of an optical trap displacement. We measured the equilibrium position of our atoms by increasing the power of the optical trap for $\approx7$ ms but not displacing it, leading to a small difference in our measured zero momentum from the actual zero momentum. These effects, up to a 20$\%$ momentum correction, were all included in our simulations.

Small fluctuations in our laboratory magnetic bias field lift the energy degeneracy of the two pseudospin states, leading to fluctuations in the pseudospin population distribution. When working with a mixture of pseudospins, we discarded any measurement for which the population of one spin state was greater than 150$\%$ of the other, resulting in up to 60$\%$ of the data from each sequence being omitted from analysis. In addition, when both dressed spins were used together, there was an initial spatial segregation of the spins due to a repulsive interaction between them\cite{Spielman2011, CZhang2012, Zhang2011}. Although the in-situ spatial distribution of the spins was modified before the experiment began, this interaction energy did not significantly affect our momentum measurements, since the in situ displacement was small compared to the typical TOF displacements giving the momentum signal.

\subsection{Simulations}
Since transitions between the dressed-spin bands are energetically suppressed due to the large energy gap between bands (compared to the energy of the dynamics), the Heisenberg equations of motion for our system were the same as Hamilton's classical equations of motion in the lowest band. In our simulation, the classical Hamiltonian included the modified position-dependent dispersion relation along \ex~(Fig.~1{\bf b}), the scalar potential from the Raman beams, the scalar potential from the optical dipole trap, and the gravitational potential. The dispersion relation was calculated by diagonalizing our system's spin-orbit coupled Hamiltonian\cite{Spielman2011} and retaining only the lowest energy band. It is the position-dependent modified dispersion relation that drives the observed SHE. The solutions to Hamilton's coupled differential equations yielded values for the position and momentum (or quasimomentum) in all three spatial directions as a function of time. For a given dressed spin, the simulated mechanical momentum $K_x$ was the difference between $q(t)$ and the location of the minimum of the dispersion curve associated with that dressed spin. Our model does not predict values of $K_y > 1.2~\hbar k_{\rm{R}} $, but this can be explained by deviations of our optical trap from the ideal Gaussian beams used in our model.

\subsection{Linear Dresselhaus and Rashba spin-orbit coupling as a vector potential}

Consider the Rashba and linear Dresselhaus spin-orbit coupling Hamiltonians in 2D\cite{Schliemann2006},
\begin{align*}
\mathcal{\check{H}}_R &= \frac{\alpha}{\hbar}\left(\hat{p}_x \check{\sigma}_2-\hat{p}_y \check{\sigma}_1\right), & {\rm and} &&
\mathcal{\check{H}}_D &= \frac{\beta}{\hbar}\left(\hat{p}_y \check{\sigma}_2-\hat{p}_x \check{\sigma}_1\right),
\end{align*}
where $\check{\sigma}_{123}$ are the Pauli spin matrices, $\hat{p}_i$ is the momentum along the $i \in \{\bm{e}_x, \bm{e}_y, \bm{e}_z\}$ spatial direction, and $\alpha$ ($\beta$) is the strength of the Rashba (Dresselhaus) SOC. The total Hamiltonian containing both of these terms,
\begin{equation*}
\mathcal{\check{H}}_{SOC} = \frac{\bm{\hat{p}}^2}{2 m} + \mathcal{H}_D + \mathcal{H}_R,
\end{equation*}
can be expressed as
\begin{equation*}
\mathcal{H}_{\rm{SOC}}=\frac{1}{2 m}\left(\check{\mathbb{1}}\bm{\hat{p}} - \bm{\mathcal{\check{A}}}\right)^2 -\frac{ m}{\hbar^2}\check{\mathbb{1}}\left(\alpha^2 + \beta^2 \right), 
\end{equation*}
with
\begin{equation}
\label{eq:soca}
 \bm{\mathcal{\check{A}}} = -\frac{m}{\hbar}  \left( \alpha \check{\sigma}_2 - \beta \check{\sigma}_1 ,  \beta \check{\sigma}_2 - \alpha \check{\sigma}_1 , 0    \right).
\end{equation}
The generalized magnetic field from this vector potential is
\begin{equation}
\label{eq:bfield2}
\bm{\mathcal{\check{B}}} = \bm{\nabla}\times\bm{\mathcal{\check{A}}}-\frac{i}{\hbar}\bm{\mathcal{\check{A}}}\times\bm{\mathcal{\check{A}}} =\bm{\mathcal{\check{B}}} = \left[\frac{2 m^2}{\hbar^3}\left(\alpha^2-\beta^2\right) \check{\sigma}_3 \right]\bm{e}_z .
\end{equation} 

\subsection{Lorentz Force}
A generalized magnetic field defined by equation (\ref{eq:bfield2}) gives a generalized Lorentz force law. Following Sakurai\cite{Sakuraibook}, we start with a Hamiltonian
\begin{equation*}
\label{eq:ham}
H = \frac{1}{2m}(\check{\mathbb{1}}\bm{\hat{p}}-\bm{\mathcal{\check{A}}})^2.
\end{equation*}
containing a non-Abelian vector potential in three spatial dimensions with a finite number of internal degrees of freedom. The Heisenberg equation of motion for the position $\bm{\hat{x}}$ is
\begin{equation*}
\label{eq:heis}
\frac{d\hat{x}_i}{dt}=\frac{1}{i \hbar}\left[\hat{x}_i, H\right] = \frac{1}{m}\left(\check{\mathbb{1}}\hat{p}_i-\check{A}_i\right) \equiv \frac{1}{m}\check{\Pi}_i.
\end{equation*}
We identify $\bm{\check{\Pi}}$ as the particle's mechanical momentum. The commutator $[\Pi_i, \Pi_j] = i\hbar \epsilon_{ijk} \check{\mathcal{B}}_k$, or $\bm{\mathcal{\check{B}}} = \bm{\nabla}\times\bm{\mathcal{\check{A}}}-\frac{i}{\hbar}\bm{\mathcal{\check{A}}}\times\bm{\mathcal{\check{A}}}  $ defines the generalized magnetic field ($\epsilon_{ijk}$ is the Levi-Civita symbol). For Abelian vector potentials, $\bm{\mathcal{\check{A}}}$ commutes along different directions, and this definition of $\bm{\check{\mathcal{B}}}$ reduces to the familiar $\bm{\mathcal{B}} = \bm{\nabla} \times \bm{\mathcal{A}}$.

We derive the Lorentz force law starting with the Heisenberg equation of motion for the mechanical momentum
\begin{equation*}
\frac{d\bm{\check{\Pi}}}{dt}=m\frac{d^2\bm{\hat{x}}}{dt^2}=\frac{1}{i \hbar} [\bm{\check{\Pi}}, H].
\end{equation*}
For an individual component
\begin{equation*}
\label{eq:lorderive}
[\check{\Pi}_i, H] =\frac{i \hbar}{2m}\left(\check{\mathcal{B}_k}\check{\Pi}_j-\check{\mathcal{B}}_j \check{\Pi}_k+\check{\Pi}_j \check{\mathcal{B}}_k - \check{\Pi}_k\check{\mathcal{B}}_j\right),
\end{equation*}
which is the $i^{th}$ component of the symmetrized Lorentz force law
\begin{equation*}
\label{eq:lorentz}
\bm{F} = m\frac{d^2\bm{\hat{x}}}{dt^2}=\frac{1}{2}\left(\frac{d\bm{\hat{x}}}{dt}   \times   \bm{\check{\mathcal{B}}}  -   \bm{\check{\mathcal{B}}} \times  \frac{d\bm{\hat{x}}}{dt}\right).
\end{equation*}

Since the $\bm{\check{\mathcal{B}}}$ field from linear combinations of Rashba and Dresselhaus SOC [equation (\ref{eq:bfield2})] and the $\bm{\check{\mathcal{B}}}$ from our experiment are both proportional to $\check{\sigma}_3$, the equation of motion for the mechanical momentum in the two cases are the same. However, for the vector potential in equation (\ref{eq:soca}), $\bm{\check{\mathcal{B}}}$ does not commute with the Hamiltonian, leading to an additional Heisenberg equation of motion for $\bm{\check{\mathcal{B}}}$ which must be included. Despite this additional complexity, the SDLF generates the SHE in both situations.

\subsection{Gauge Invariance}

The magnetic field defined by equation (\ref{eq:bfield2}) is not gauge-invariant. The definition of gauge transformations is generalized in any discussion of non-Abelian vector potentials. For the SU(2) symmetry group, a gauge transform is a position-dependent unitary rotation in spin-space\cite{Schroederbook, Mills1954}
\begin{align*}
\psi &\rightarrow \check{V}(\bm{\hat{x}}) \psi, &{\rm with} && \check{V}(\bm{\hat{x}})&=\rm{exp}[ \bm{\alpha}({\bf \hat{x}}) \cdot \bm{\check{\sigma}}],
\end{align*}
where $\bm{\alpha}$ is an arbitrary vector of functions of ${\bf \hat{x}}$ and $\bm{\check{\sigma}}$ is the vector of 2 $\times$ 2 Pauli matrices including the identity. Under this gauge transformation, the Lagrangian must remain unchanged, requiring the magnetic field to transform according to\cite{Schoutens2011}
\begin{equation*}
\bm{\mathcal{\check{B}}} \rightarrow \check{V}(\bm{\hat{x}})\bm{\mathcal{\check{B}}}\check{V}^\dagger(\bm{\hat{x}}).
\end{equation*}
Despite the lack of gauge invariance of the magnetic field, an Abelian magnetic field cannot be gauge transformed to a non-Abelian field. 

This definition for gauge transforms can be generalized to a gauge with generators from any continuous symmetry group by replacing $\bm{\check{\sigma}}$ with a vector of the generators of the symmetry group. For instance, in the case of a scalar vector potential from classical electrodynamics with $U(1)$ symmetry, the generator of the symmetry group is a scalar, and the gauge transformation becomes the familiar position-dependent phase.

\end{methods}

\begin{suppmat}

{\bf Introduction} In the Methods, we argued on the basis of our simple conceptual picture of equal and opposite synthetic magnetic fields that our scheme can realize a quantum spin Hall effect (QSHE).  Here we make this argument rigorous by solving the full Raman-coupled Hamiltonian for a 2D fermionic system that is extended along \ex; has a Raman coupling profile $\Omega(y)$; and is confined in a potential $V_{\rm tot}(y) = V_{\rm box}(y) + V_{\rm comp}(y)$ along \ey, where the two terms are a box potential
\begin{align*}
V_{\rm box}(y) &= \Bigg\{
\begin{array}{cl}0 & \text{for $y\in\left(0,L_y\right)$} \\\infty & \text{otherwise}\end{array}, 
\end{align*}
along with a compensation potential $V_{\rm comp}(y)$ described below. $L_x$ and $L_y$ describe the system's extent along \ex~and \ey.  Assuming that the atoms remain everywhere in the lowest band of dressed states, the stationary Schr\"odinger equation is
\begin{align}
E\psi(x,y) &= \left[\mathcal{E}_-\left(-i \frac{\partial}{\partial x},\Omega(y)\right) - \frac{\hbar^2}{2 m}\frac{\partial^2}{\partial y^2} + V_{\rm tot}(y)\right] \psi(x,y),
\end{align}
where $\mathcal{E}_-(\hat{q},\Omega)$ are energies of the ground-band Raman dressed states (the double well pictured in Fig.~1 of the manuscript).  As with the solution to the case of an electron moving in a uniform magnetic field expressed in the Landau gauge, this problem can be solved by taking a separable wavefunction of the form $\psi(x,y) = \exp(i q x) f_{q}(y)$.  With this ansatz, the Hamiltonian reduces to the 1D problem
\begin{align}
E_{q} f_{q}(y) &= \left\{- \frac{\hbar^2}{2 m}\frac{d^2}{d y^2} +\bigg[ V_{\rm tot}(y) +\mathcal{E}_-\left( q,\Omega(y)\right)\bigg]\right\} f_{q}(y), \label{eq:Hamiltonian1DSpinless}
\end{align}
with an additional $q$-dependent ``potential'; this would be harmonic for a charged particle in a uniform magnetic field.  Here, $\mathcal{E}_-(q,\Omega(y))$ contributes an additional scalar potential dependent on $\Omega$:  as $\Omega$ increases and the minima of the two wells move together, they also move downwards in energy by $-\left[\hbar\Omega(y)\right]^2/16\Er$.  A compensation potential $V_{\rm comp}(y)$ is added to cancel this term.  For non-trivial profiles $\Omega(y)$, the resulting  $V_{\rm comp}(y)$ is also complicated, but in the specific proposal described below, we see that the conventional gaussian laser profiles suffice for both.

{\bf Full model} While the approximation that atoms remain in the ground band provides a useful conceptual framework for discussing this system, we can solve the full two level problem with equal ease.  Following the same reasoning leading to Eq.~(\ref{eq:Hamiltonian1DSpinless}), the full 1D spinful Hamiltonian describing motion along \ey~is
\begin{align}
E_{q} \chi_q(y) &= \left\{\left[- \frac{\hbar^2}{2 m}\frac{d^2}{d y^2} + \frac{\hbar^2\left(q^2+k_R^2\right)}{2m} + V_{\rm tot}(y)\right]\check{1} + \frac{\hbar\Omega(y)}{2}\check\sigma_1 - \frac{\hbar^2 k_R q}{m}\check\sigma_3 \right\}\chi_q(y),\label{eq:Hamiltonian1DSpin}
\end{align}
where $\chi_q(y) = \left\{f^\uparrow_q(y),f^\downarrow_q(y)\right\}$ is a two component wavefunction; $\check{1}$ is the $2\times2$ identity; and $\check\sigma_{1,2,3}$ are the Pauli matrices.

In the Methods, our conceptual example used a carefully selected $\Omega(y)$ giving a spatially homogeneous spin-dependent magnetic field with magnitude $\mathcal{B}_0$.  This conceptual nicety is not important for realizing the QSHE; the simple gaussian profiles
\begin{align*}
\hbar\Omega(y) &= 4\Er e^{-2\left(y/w_0\right)^2}, &{\rm and} && V_{\rm comp}(y) &= \alpha \Er e^{-4\left(y/w_0\right)^2}
\end{align*}
are sufficient to create a QSHE system.  $w_0$ is the $1/e^2$ radius of the Raman lasers and $\alpha$ sets the scale of the compensation potential ($\alpha = 1$ is the compensation predicted by considering only the energy of the band minima).  In practice, we find a robust QSHE for $w_0\approx L_y$  (optimal for $w_0 = 1.15 L_y$) and $\alpha \approx 1$ ($\alpha = 0.995$ leads to a slightly more uniform gap).  The proposed geometry is depicted in Fig.~\ref{fig:QSHEbands}{\bf a}.  In practice, a separate compensation laser is not required, and the ac Stark shift from the Raman laser beams suffices.  In this case, one selects one Raman beam to have a much larger beam waist than the other; recalling that the Raman coupling $\Omega\propto \sqrt{I_1 I_2}$ (where $I_{1,2}$ are the beam intensities), this implies that $\Omega$ is largely shaped by the profile of the smaller beam, to the $1/2$-power.  In contrast, the ac Stark shift is still given by the intensity, and therefore scales like $\Omega^2$, as required.

The four smallest eigenenergies of this Hamiltonian are plotted as a function of $q$ in Fig.~\ref{fig:QSHEbands}{\bf b}, showing that each state is specified by two quantum numbers, $q$ and a Landau-level-like index $N$.  We numerically verified that the low-energy spectra of the approximate [Eq.~(\ref{eq:Hamiltonian1DSpinless})] and exact [Eq.~(\ref{eq:Hamiltonian1DSpin})] Hamiltonians are indistinguishable.  This is because atoms always reside near the local minima in $\mathcal{E}_-(q,\Omega(y))$.  Near these minima, the band spacing $\mathcal{E}_+(q,\Omega(y)) - \mathcal{E}_-(q,\Omega(y))>4\Er$ even as $\Omega\rightarrow0$, large compared to the $200\Hz\approx0.05\Er$ interval between eigenenergies depicted in Fig.~\ref{fig:QSHEbands}{\bf b}.

For very small Fermi energy, $E_{{\rm F},1}$ (pale horizontal line in Fig.~\ref{fig:QSHEbands}{\bf b}), the full SHE analogy is revealed, with four points at the Fermi energy, corresponding to the depicted edge modes on the top and bottom of the system (Fig.~\ref{fig:QSHEbands}{\bf b}, empty circles).  Figure~\ref{fig:QSHEbands}{\bf c} (dashed curves) plots the computed density distribution of those eigenstates, showing that they reside on the system's edge, and on a given edge, the two spins' edge states counterpropagate with opposite group velocities (direction indicated by the sign of the density curves).

However, for the larger (and more practical) Fermi energy $E_{{\rm F},2}$, the two edge states at the system's bottom (with $q\approx0$) hybridize as the two minima comprising our pseudo-spins merge, while the edge states at the top remain robust (solid red and blue curves in Fig.~\ref{fig:QSHEbands}{\bf c}).  Earlier, we made analogy to the most simple model QSHE system which consists of a pair of superimposed IQHE systems, one for each spin, with equal and opposite magnetic field.  The vanishing edge states on one boundary of our system have a similar analog. In this case, we consider a single IQHE system in which the microscopic spin smoothly twists from up to down while moving across the system along \ey.  If such a system is folded so that the original bottom edge overlaps with the top edge, the resulting system has overlapping spin-dependent edge states on the top side and not on the bottom (where the fold is); the apparent edge on the folded side is illusory.  

{\bf Physical parameters} To estimate the number of fermions required, we consider a uniform system with length $L_x$, constraining $q$ to be a multiple of $\delta q = 2\pi/L_x$.  Here the edge states are well-isolated for a Fermi momentum $k_{\rm F} \approx k_{\rm{R}} = 2 \pi / \lambda$ (solid red and blue circles in Fig.~\ref{fig:QSHEbands}b). Since this energy is fully in the gap, it implies that states with $\left|q\right| \leq k_{\rm F}$ are occupied.  Thus, the number of states below the Fermi energy is $N_F = 2 L_x / \lambda$.  For example, when $L_x = 50\lambda$ the atom number will be $N_{\rm F} = 100$, and the proposed system with height $L_y = 7.75\lambda$ has a reasonable 6:1 aspect ratio.  With these $\approx 100$ atoms, the Fermi energy will lie in the gap between the $N=0$ and $N=1$ bands, which are spaced by $\approx 200\ {\rm Hz} \approx 10\ {\rm nK}$.  As with existing proposals for creating topological matter with cold atoms\cite{Jaksch2003, Osterloh2005, Ruseckas2005, Gunter2009, Spielman2010, Cooper2011a,  Cooper2011, Eckardt2012, Chen2012}, the required energy scales and atom numbers are low, but within the scales that have already been realized in the lab\cite{Esslinger2008,Jochim2011}.  In this case, to identify the existence of edge states, a recent technique proposed by Goldman {\it et. al.}\cite{Spielman2012a} could be implemented to directly image the spin-dependent motion of edge states.

{\bf Conclusion} This proposal requires: (1) a box potential along \ey, e.g., Ref.~\onlinecite{Hadzibabic2012}, (2) a conventional harmonic potential along \ex, (3) gaussian Raman laser beams, (4) an anti-confining gaussian laser compensating for the $\Omega(y)$ dependence of the band-minima along \ey; and (5) confinement in 2D in the $x-y$ plane. This novel method for creating the QSHE has advantages over previous proposals for creating the QSHE in a quantum gas\cite{Jaksch2003, Osterloh2005, Ruseckas2005, Gunter2009, Spielman2010, Cooper2011a,  Cooper2011, Eckardt2012, Chen2012}. Our conceptually simple method for designing the vector potential relies on only two lasers and two atomic states, making the experimental implementation relatively easy. One of the major stumbling blocks in using alkali atoms for the QSHE is the large photon scattering rate from the Raman lasers for alkali fermions, which leads to significant heating of the sample\cite{Spielman2010}. Our approach here ameliorates this concern by using low coupling strengths ($\hbar \Omega < 4~E_{\rm{R}}$ over the entire sample), requiring less laser intensity.


\begin{figure}[t!]
\begin{center}
\includegraphics[scale=1]{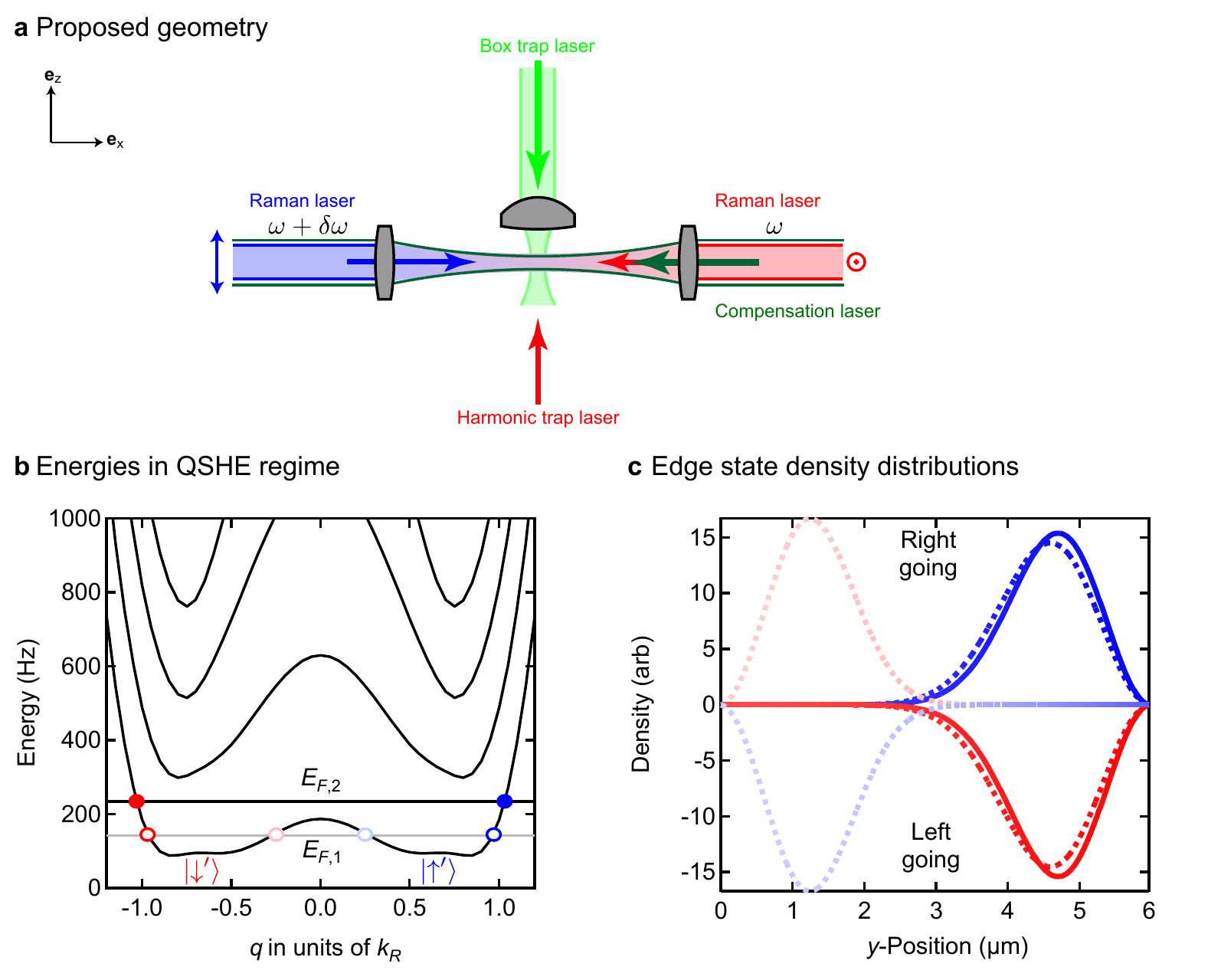}
\end{center}
\caption{{\bf Quantum spin Hall effect.} {\bf a} Side view of laser geometry including: $\lambda\approx770\nm$ Raman lasers; $\lambda\approx532\nm$ compensation lasers and box lasers; and a $\lambda\approx1064\nm$ harmonic trap laser for confinement along ${\bf e}_x$. {\bf b} Energies for QSHE geometry using $\hbar \Omega(y) = 4 E_{\rm{R}} \exp\left[- 2\left (y/ w_0 \right)^2 \right]$, and $V_{\rm comp} = \alpha E_{\rm{R}} \exp\left[- 4\left (y/w_0 \right)^2 \right]$, where $L_y = 7.75\lambda$, $w_0 = 1.15 L_y$, and $\alpha = 0.995$.  For $E_{F,1} = 142\Hz$, the empty red and blue symbols at $q = \pm 0.97\kr$ and the empty light-red and light-blue symbols at $q = \pm 0.25\kr$ mark the points at the Fermi surface.  For $E_{F,2} = 233\Hz$, the solid red and blue symbols at $q = \pm 1.03\kr$ mark the states at the Fermi surface.  {\bf c} Density distributions corresponding to the states at the six circles marked in {\bf b}, color-matched to the circles' fill colors, and where the dashed lines correspond to the empty symbols.  The sign of the density distributions indicate the direction of propagation for that state, corresponding the the group velocity in {\bf b}. 
\label{fig:QSHEbands}}
\end{figure}

\end{suppmat}

\end{document}